\documentstyle[aps,prb,preprint]{revtex}
\begin{document}
\draft
\title{Spectral properties of the $t-J$ model in the presence
       of hole-phonon interaction}
\author{Bumsoo Kyung$^{a}$,  Sergei I. Mukhin$^{b}$,
        Vladimir N. Kostur$^{c}$, and Richard A. Ferrell$^{a}$}
\address{$^{a}$Department of Physics, Center for Superconductivity Research, 
University of Maryland, \\ College Park, MD 20742-4111\\
$^{b}$Moscow Institute of Steel and Alloys,
Theoretical Physics Department, Leninskii pr. 4, 117936\\ Moscow B-49,
Russia\\
$^{c}$Department of Physics,
S.U.N.Y. at Stony Brook, Stony Brook, NY 11794-3800}

\maketitle
\begin{abstract}

We examine the effects  of electron-phonon interaction on the dynamics
of the charge carriers doped in two-dimensional (2D) Heisenberg 
antiferromagnet. 
The $t$-$J$ model Hamiltonian with a Fr\"ohlich term which couples the holes
to a dispersionless (optical) phonon mode is considered for  
low doping concentration.
The evolution of the spectral density function, the density of states,
and the momentum distribution function of the holes with an increase of the 
hole-phonon coupling constant $g$ is studied numerically.
As the coupling to a phonon mode increases the quasiparticle spectral weight
decreases and a ``phonon satellite'' feature close to the quasi-particle
peak becomes more pronounced. Furthermore, strong electron-phonon coupling
smears the multi-magnon resonances (``string states'') in the incoherent
part of the spectral function. The jump in the momentum distribution function
at the Fermi surface
is reduced without changing the hole pocket volume,
thereby providing
a numerical verification of  Luttinger theorem for this strongly
interacting system. The vertex corrections due to electron-phonon interaction
are negligible  
in spite of the fact that the ratio of 
the phonon frequency to the effective  bandwidth
is not small.

\end{abstract}
%
%%%%%%%%%%%%%%%%%%%%%%%%%%%%%%
%
%
\section{INTRODUCTION}
\label{section1}
Since the discovery of high-temperature superconductivity in copper-oxide 
based materials, experimental studies have revealed a lot of evidence that
interaction between electrons and lattice vibrations 
plays an important role in these compounds.\cite{Lattice} The changes
in position and  width of the phonon peaks  
below superconducting transition temperature measured by Raman 
spectroscopy\cite{Thomsen} and by neutron scattering techniques\cite{Mook}, 
the existence of the isotope effect  which varies 
as a function of both doping and rare earth ion substitution\cite{Franck}, 
clearly demonstrate
that coupling between charge carriers and phonon modes in high-$T_c$ superconductors
is not negligible. 
Nevertheless, theoretical studies of the effects of electron-phonon
interaction on the dynamics of the charge carriers in  electron-correlated
systems are 
far from complete. The main problem for proper treatment of 
electron-phonon interaction in high-$T_c$ superconductors
is that charge carrier motion in an antiferromagnetic background (or spin-fluctuating 
background at higher doping concentration) 
is strongly affected by electron-magnon interactions itself.
\cite{Auerbach,Schmitt-Rink:1988,Kane,Martinez:1991,Igarashi,Liu:1992}
Including the electron-phonon interaction deals with quasiparticles renormalized
by  interaction with magnons.\cite{Ramsak:1992}
 As a consequence the renormalized  bandwidth becomes comparable
to phonon frequencies and  the `classical' Migdal-Eliashberg approach
to the  electron-phonon
problem\cite{Migdal:1958}
 in metallic systems seems  beyond the region of application.

   In the present paper, we add a Fr\"{o}hlich term to the $t-J$ model 
Hamiltonian to study 
the quasiparticle properties in the presence of hole-phonon
interaction.
This is particularly interesting, since it is widely believed that
the $t-J$ model incorporates essential features of 
electron systems with strong local repulsion between electrons, 
characteristic of high T$_{c}$ copper oxides.\cite{Anderson} 
We examine the properties of this model 
 at low doping concentrations (near half-filling)
where analytical and numerical results for the $t-J$ model are
well-known.
The optical phonon mode in the 
copper-oxide plane is assumed to have a substantial influence 
on the dynamics of
doped holes in cuprate compounds.
In the past, quasiparticle properties 
in the presence of electron-phonon interaction 
were studied by Engelsberg and Schrieffer \cite{Engelsberg:1963} 
for a weakly correlated model of conventional metal.
The authors 
examined the spectral density function $A(\vec{k},\omega)$ of an 
electron
for phonon spectra of the Einstein and Debye forms. 
They found that spectral density function 
exhibits several branches of
excitations rather than a single branch of a dressed electron. 
Our calculations of the quasiparticle energy dispersion 
show the same
for the present model of a  strongly correlated 
system with hole-phonon interaction 
 \cite{Kyung:19961}.
Recently the mass renormalization
due to a coupling to optical phonons was studied for strongly
correlated electrons by Ram\u{s}ak {\em et al.} 
\cite{Ramsak:1992}. 
The authors found that the phonon-induced mass
renormalization of a single hole that propagates in the $t-J$ model on
 the
 scale $2J$ is much larger than that in the corresponding uncorrelated model
for $J \agt t$.  
The mass enhancement, increasing with $J/t$ ratio, is due to the slow 
motion of a spin
polaron, which makes hole-phonon interaction more effective.
The hole excursions
from the center of the spin polaron are restricted to a few lattice
sites and the bandwidth is on scale $t^{2}/J$, as long as $J/t$ is large. 
On the other hand, when $J\ll t$  
the confining
antiferromagnetic potential becomes weak and a hole performs large radius
incoherent excursions on scale $t$. Therefore, the mass enhancement 
induced by the hole-phonon 
interaction becomes weaker and the coherent part of the spectral weight 
tends to zero.

In our paper,   we study 
various quasiparticle properties as well as
the mass renormalization due to hole-phonon
interaction in the limit of a finite but small concentration of doped holes.
The experimentally
measured optical phonon frequencies \cite{Falck:1993}
are close to 50 $meV$ and typical values of $J$ are
of the order of 100 $meV$.
We assume the optical phonon frequency, $\Omega$,
to be  $\Omega=0.5J$. 
The $t-J$ model Hamiltonian appended with Fr\"ohlich term 
will be denoted as the $t-J-g$ model.
Our paper is organized as follows.
In Sec.~\ref{section2}, the quasiparticle residue and the mass
renormalization constant are calculated for $t-J-g$ model. 
The incoherent part of the spectrum and Luttinger's theorem are studied in 
Sec.~\ref{section3}.
Vertex corrections due to the hole-phonon interaction are analyzed 
in Sec.~\ref{section4}.
Section \ref{section5} presents the optical conductivity in the 
$t-J-g$ model. 
A summary is given in Sec.~VI.
\section{FORMULATION AND QUASI-PARTICLE SPECTRAL WEIGHT}
\label{section2}

   To analyze the interaction between optical phonons
and charge carriers in the copper oxide planes, we consider a two-dimensional 
$t-J$ model Hamiltonian appended by a Fr\"{o}hlich term \cite{Mahan:1986}
with the dispersionless (optical)
phonon mode $\Omega$ 
and  the coupling constant $g$. Using linear spin-wave approximation through
Holstein-Primakoff transformation\cite{Primakoff} and spinless-fermion/
Schwinger-boson representation\cite{Auerbach} of the electron operators
on a site $i$ with spin $\sigma$ 
$c_{i\sigma}$: $c_{i\downarrow}^{\dagger} \rightarrow h_i S_i^-$ and
$c_{i\uparrow} \rightarrow h_i^{\dagger}$ the Hamiltonian can be written:

\begin{eqnarray}
  H_{tJg} & = & -t\sum_{\langle i,j\rangle}\left[h_{i}h^{+}_{j}S^{-}_{j}+
  S^{-}_{i}h_{i}h^{+}_{j}+h.c.\right]
  +J\sum_{\langle i,j\rangle}\rho^{e}_{i}\vec{S}_{i}\cdot
  \vec{S}_{j}\rho^{e}_{j}
\nonumber \\
  & - & \mu\sum_{i}\rho^{h}_{i}+
  g\sum_{i}h^{+}_{i}h_{i}(b_{i} + b^{+}_{i})+
  \Omega\sum_{i}b^{+}_{i}b_{i} \; ,           
                                                         \label{eq:226}
\end{eqnarray}
where $h_i$ is a hole creation operator on site $i$ and $\vec{S}_{i}$   
is the on-site electron spin operator.
($S^{-}_{i}$ turns down the spin of the electron on site $i$.)
The summation is over the nearest neighbor 
sites $\langle i,j\rangle$ of 
the square
lattice, $b^{+}_{i}$ is an optical phonon creation operator at site $i$.
$\rho^{h}_{i}\equiv h_{i}^{+}h_{i}$ is the local density 
operator of the spinless
hole, while $\rho^{e}_{i}=1-\rho^{h}_{i}$ is the local
density operator of electron
under the single occupancy constraint. 
Further, $\mu$ is 
the chemical 
potential related to the concentration of the doped holes:
$\langle\rho^{h}_{i}\rangle = x$ (at half-filling $x=0$). 

   Since $t>J$ would correspond to a situation in cuprates, 
the usual perturbation 
approximation does not work in the strongly coupled $t-J$ model.
However, it has been for some time
known that the self-consistent noncrossing approximation for 
the hole self-energy gives a fairly accurate result because of the 
small vertex corrections. Namely, it was explained in Ref.~\cite{Sush},
that the leading correction to the hole-spin wave vertex 
vanishes as a consequence of
the electron (hole) spin conservation in the $t-J$ model. 
Mathematically, this fact
is reflected in the special symmetry of the hole-spin wave interaction 
vertices 
$M_{\vec{k},\vec{q}}$ and $N_{\vec{k},\vec{q}}$, which leads to the identical 
vanishing of non-maximally crossing hole-spin wave vertex corrections 
\cite{Liu:1992}.
However, it is uncertain whether or not the vertex corrections due to
the hole-phonon interaction are negligible in a strongly correlated 
electron system. Thus, in the first place we assume that they are so, and 
later in Sec.~\ref{section4} this assumption will be 
supported by calculating the 
lowest order vertex corrections.
To study the influence of
hole-phonon coupling on the quasiparticle properties we treat the
Hamiltonian within the noncrossing approximation
for spin wave and phonon interactions 
\cite{Ramsak:1992}.
Fig.~\ref{fig110} shows
four possible noncrossing diagrams contributing to the self-energy:
spin wave emitting and absorbing diagrams, and 
phonon emitting and absorbing ones.
The thick solid line is
the dressed hole Green's function, while 
the unperturbed hole propagator is denoted by the thin solid line.
The dashed and wavy lines 
stand for the 
 spin wave and phonon propagators, respectively.
Here $M$, $N$ are the hole-spin wave interaction vertices, while $g$ is 
the hole-phonon vertex.
After summing over the intermediate frequency,
the hole self-energy at zero temperature becomes
\begin{eqnarray}
 \Sigma^{R}(\vec{k},\omega)  =  \sum_{\vec{q}} \{ M_{\vec{k},   
            \vec{q}}^{2}\int_{0}^{\infty}dy\frac{A(\vec{k}-\vec{q},y)}   
            {\omega-y-E_{\vec{q}}+i\delta}   
         + N_{\vec{k},\vec{q}}^{2}\int_{-\infty}^{0}dy\frac{A(   
           \vec{k}-\vec{q},y)}{\omega-y+E_{\vec{q}}+i\delta}   
                                                        \nonumber    \\
         +\ \  g^{2}\int_{0}^{\infty}dy\frac{A(\vec{k}-\vec{q},y)}{   
           \omega-y-\Omega+i\delta}   
         + g^{2}\int_{-\infty}^{0}dy\frac{A(\vec{k}-\vec{q},y)}{   
           \omega-y+\Omega+i\delta} \} \; ,                  \label{eq:236}
\end{eqnarray}
where the superscript $R$ indicates retarded functions which are 
analytic in the upper
half-plane of $\omega$, $E_{\vec{q}}$ is the spin wave energy, and 
$A(\vec{k},\omega)$ is the 
spectral density function of the interacting hole propagator given as
\begin{equation}
A(\vec{k},\omega) \; = \; -\frac{1}{\pi}ImG^{R}(\vec{k},\omega) \; .
                                                         \label{eq:234}
\end{equation}
The hole Green's function is then obtained from the self-energy
\begin{equation}
G^{R}(\vec{k},\omega) \; = \; \frac{1}{\omega-\Sigma^{R}(\vec{k},
                            \omega)+\mu+i\delta} \; .    
                                                          \label{eq:235}
\end{equation}
The  self-consistent integral equation, Eq.~\ref{eq:236}, is solved
numerically (See Ref.~\cite{Kyung:19961} for 
detail).
    Table~\ref{table2} shows the quasiparticle residues at three different
$\vec{k}$ points and the energy shift due to both the hole-spin wave and
hole-phonon interactions.
$a_{(\pi/2,\pi/2)}^{calc}$ in the fifth column is the quasiparticle residue 
at the point $(\pi/2,\pi/2)$ calculated
by perturbation theory. 
The numerically computed mass renormalization factor due to 
the hole-phonon interaction is $\lambda_{num}$. 
As expected, the quasiparticle residue decreases as the coupling constant
increases. At the $(0,0)$ point the hole loses its quasiparticle property more
rapidly, as
the vanishingly small spectral weight indicates. This is because 
the state at the $(0,0)$ point is located at a high hole energy region 
and hence it
is more vulnerable to  additional decay from the hole-phonon
interaction.
The energy shift
due to the hole-phonon interaction is generally small ($-0.975$ for
$g=2.0J$), compared with that 
from the pure hole-spin wave interaction $(-5.378)$. 
This suggests that a perturbative
treatment for the hole-phonon interaction is possible.
According to Eq.~\ref{eq:236}, the hole self-energy is
composed of two terms, one from the hole-spin wave interaction and 
the other from the hole-phonon interaction. Thus the self-energy can be written
as $\Sigma(\vec{k},\omega)=\Sigma_{m}(\vec{k},\omega)
   +\Sigma_{g}(\vec{k},\omega)$, where the former comes from the 
hole-spin wave interaction and the latter from the hole-phonon interaction.
If we define $X$ and $Y$ as follows
\begin{eqnarray*}
    X & = & 1 - \frac{\partial}{\partial \omega}
                  Re\Sigma_{m}(\vec{k},\omega){\mid}_{g=0} \; ,
                                                                   \\
    Y & = & - \frac{\partial}{\partial \omega}
                  Re\left(\Sigma_{g}(\vec{k},\omega)+
                  \Sigma_{m}(\vec{k},\omega)-
                  \Sigma_{m}(\vec{k},\omega){\mid}_{g=0}\right) \; ,
\end{eqnarray*}
the quasiparticle residue can be approximated by
\begin{eqnarray}
 a_{\vec{k}} & = & \frac{1}{X+Y} 
          =  \frac{1}{X}[1-(\frac{Y}{X})+(\frac{Y}{X})^{2}+\cdots]
                                                        \nonumber       \\
       & \approx & \frac{1}{X}[1-(\frac{Y}{X})+(\frac{Y}{X})^{2}] \; .
                                                      \label{eq:522}
\end{eqnarray}
The $X$, $Y$ are determined by 
\begin{eqnarray*}
    X & = & \frac{1}{a_{\vec{k}}(g=0)} \; , 
                                                                     \\
    Y & = & \frac{1}{a_{\vec{k}}} - \frac{1}{a_{\vec{k}}(g=0)} \; .
\end{eqnarray*}
Substituting $X$ and $Y$ into Eq.~\ref{eq:522} leads to the 
$a_{(\pi/2,\pi/2)}^{calc}$ in the fifth column of Table~\ref{table2}.
For coupling constants less than $1.5J$,  agreement with the numerical
results is quite satisfactory. However, the perturbative
treatment based on
the two term expansion breaks down for $g>1.5J$, since $(Y/X)$ 
increases significantly for the strong hole-phonon interaction.
In fact, $Y$ is the approximate hole-phonon mass renormalization constant 
$\lambda$ evaluated numerically on the basis of the $t-J$ model Hamiltonian.
This is listed as $\lambda_{num}$ in the sixth column.
The mass renormalization constant for a small coupling constant
can also be computed from perturbation theory.
We used a similar method employed by 
Ram\u{s}ak {\em et al.} 
\cite{Ramsak:1992}. 
Since the hole-phonon coupling constant is small compared with the 
hole-spin wave interaction strength, the effective mass of the hole can
be computed using the lowest-order perturbation correction to the
hole self-energy,
\begin{equation}
  \omega_{\vec{k}}(g)-\omega_{\vec{k}}=\frac{1}{(2\pi)^{2}}\int
          d^{2}q\frac{a_{\vec{k}-\vec{q}}g^{2}}
          {\omega_{\vec{k}}-\omega_{\vec{k}-\vec{q}}
           -\Omega} \; ,
                                                           \label{eq:523}
\end{equation}
where $\omega_{\vec{k}}(g)$ and $\omega_{\vec{k}}$ are the quasiparticle
dispersion functions in the presence of the hole-phonon interaction and
in its absence, respectively. Approximately, $\omega_{\vec{k}}$ 
is given by 
\[ \omega_{\vec{k}} = \frac{(\vec{k}-(\pm\pi/2,\pm\pi/2))^{2}}
                                                     {2m_{eff}} \; , \]
where $m_{eff}$ is the averaged effective mass of the hole near the bottom of 
the energy dispersion function, which is taken as $3.36/t$ according to  
Martinez {\em et al.} \cite{Martinez:1991}. From the second derivative 
around the points
$\vec{k}=(\pm\pi/2,\pm\pi/2)$, we arrive at 
\begin{eqnarray}
  \frac{1}{m_{eff}(g)}-\frac{1}{m_{eff}} & \approx &
            -\frac{16\bar{a}g^{2}m_{eff}}{\pi}\int_{0}^{\pi}dq
             \frac{q^{3}}{(q^{2}+2m_{eff}\Omega)^{3}}
                                                        \nonumber  \\
         & = & -\frac{1}{m_{eff}} 
             \frac{2\bar{a}g^{2}m_{eff}}{\pi\Omega}
             [4\int_{0}^{y_{m}}dy
             \frac{y^{3}}{(y^{2}+1)^{3}}] \ ,
                                                        \label{eq:524}
\end{eqnarray}
where $y_{m}$ is given by $\pi/\sqrt{2m_{eff}\Omega}$ and 
$\bar{a}$ is the averaged quasiparticle residue near the point 
$(\pi/2,\pi/2)$.
Hence,
\begin{eqnarray}
   \lambda_{eff} & = & \frac{m_{eff}(g)}{m_{eff}} - 1
    \approx  \frac{1}{1-1.5\bar{a}g^{2}m_{eff}/(\pi\Omega)} - 1
                                                        \nonumber  \\
   & \approx & 1.5\bar{a}g^{2}m_{eff}/(\pi\Omega) \; ,
                                                        \label{eq:5241}
\end{eqnarray}
where a small $\lambda_{eff}$ is assumed.
For $g/J = 0.5$ and 1.0, the substitution of $\bar{a} \approx 0.34$
gives 0.127 and 0.822, respectively. 
The former value is favorably compared with 0.102, the numerically 
calculated mass renormalization constant for $g/J = 0.5$.
Clearly $g/J = 1.0$ is too strong to apply perturbation theory,
because of the too large effective mass, $m_{eff}=3.36/t$. 
Besides which, the large value of 
$\sqrt{2m_{eff}\Omega}$, namely, 1.16 makes the calculation
very sensitive to the 
upper bound, $y_{m}$, for the integration, leading to an additional difficulty.
A straightforward calculation of $\lambda_{eff}$ for noninteracting 
electrons yields 0.016, 0.068, 0.167 and 0.342 for $g/J=0.5$,
1.0, 1.5 and 2.0, respectively. In this case, the above mentioned
difficulties do not occur because of the small effective mass,
$m_{eff}=1/2t$.
The calculation shows that the mass renormalization factor is 
more enhanced 
for the strongly correlated electrons
than the factor for the noninteracting
electron system. This enhancement factor which is 6.4 for $g/J$ = 0.5
 is substantially large, compared
with 3.5 which
Ram\u{s}ak {\em et al.} \cite{Ramsak:1992} reported.
The discrepancy between these two values originates from the 
use of different definitions for the mass enhancement parameter.
The authors of this paper obtained the enhancement factor
by explicitly computing the change in the curvature of the quasiparticle
dispersion
along the line $(\pi/2,\pi/2)-(0,0)$, namely, $\lambda_{\parallel}= 
m_{\parallel}(g)/m_{\parallel}-1$, while the effective
mass in the present calculation, 
$\lambda_{eff}=\sqrt{
m_{\parallel}(g)m_{\perp}(g)/m_{\parallel}m_{\perp}}-1$,
is averaged out
around the point $(\pi/2,\pi/2)$.
This indicates even larger mass enhancement along the line
$(0,\pi)-(0,0)$, which is consistent with the observation 
that the hole moves slowest along this direction.
\section{INCOHERENT SPECTRUM AND LUTTINGER'S 
            THEOREM} 
\label{section3}

   Because of the additional scattering channel, we expect that  
there are more spectra associated with the optical phonon excitations 
above the quasiparticle pole in the spectral density function.
Fig.~\ref{fig111} shows the spectral density function for four different
hole-phonon coupling constants $g$.
As $g$ increases, the spin wave peaks become  more suppressed, whereas the
phonon peaks become stronger.   
Since the coupling strength for the hole-spin wave interaction is much
larger than that for the hole-phonon interaction, the weak phonon
features appear on top of the spin wave peaks as a satellite structure.  
For $g \geq 1.5J$, the phonon induced peaks 
are even sharper and larger in height than the 
spin wave peaks. Especially the appearance of the multiple
phonon peaks below the chemical potential for $g=2.0J$, compared 
with the spectral density function for $g=0$, is clearly noticeable.
The spikes due to finite size effects are gradually
reduced, since those artificial peaks get smeared out due to the
additional decay induced by the hole-phonon interaction.
Fig.~\ref{fig112} presents the hole density of states for various
coupling constants.
The general behavior for the hole density of states is similar to that for the 
spectral density function, since in most of the Brillouin zone the
spectral density function is very similar to that at the point ($\pi/2,\pi/2$).
As the hole-phonon coupling constant increases, the position of the  
strongest phonon peak right above the Fermi energy ($\omega = 0$) 
is shifted upward, although this is not well pronounced due to finite
size effects.
This is understood based on a (lattice) polaronic formation. 
For the strong hole-phonon coupling constant, the 
hole is surrounded by an increasing number of phonons.  
Hence, approximately 2 phonons are involved in the formation of the polaron
for $g=2.0J$, since the position of the peak is close to $1.0J$ and
the optical phonon frequency is $0.5J$.
The figure for $g=0$ shows, however, 2.5 spin waves with energy $2J$ 
(i.e. from the top of the spin wave band where the density of spin wave
states is sharply peaked) participate in a magnetic polaron.
The incoherent spectrum below the Fermi energy in the density of states
is crucial in satisfying Luttinger's theorem 
\cite{Kyung:19961}, as far as the quasiparticle (coherent) contribution to the
density of the occupied states (at $T=0$) is substantially suppressed due to
the strong hole-spin wave (and phonon) coupling.
The momentum distribution function is shown in Fig.~\ref{fig113}.
We chose a much larger cluster ($240 \times 240$) for $n(\vec k)$ calculation, 
than the original cluster ($24 \times 24$) used for
the self-consistent calculation
of the self-energy in the $\vec{k}$ points along 
the line $(\pi/2,\pi/2)-(0,\pi)$ or
$S-Y$. This provides detailed information about the
behavior of $n(\vec{k})$ in the
vicinity of the Fermi surface. 
The four very elongated ellipses in the inset denote the Fermi surface.
The distribution function shows a sharp drop  
{\em at the same $\vec{k}$ point} for all the coupling constants we have
studied, as seen in the figure.
We also compared the doping concentration $x$ computed from 
the spectral density function, with 
the ratio of the number of $\vec{k}$ states inside the Fermi surface 
to the number in the entire (antiferromagnetic) Brillouin zone.
The former yields $x=0.030$, 0.031, 0.032, 0.030, 0.030 for 
$g=2.0J$, $1.5J$, $1.0J$, $0.5J$, $0$, respectively, while 
the latter shows 0.032. They agree with each other within less than
$4 \%$ on the average.
These two features numerically verify Luttinger's theorem
in the $t-J-g$ model for
a small doping concentration.
As the coupling constant increases, the distribution function inside the
Fermi surface decreases gradually from 0.36 to 0.23.
This is due to the reduced quasiparticle residue for the strong
hole-phonon interaction, as seen in Table~\ref{table2}.
At the same time, however,
some density of the occupied states also appears outside the Fermi surface.  
This is because in the $t-J-g$
model the spectral density function $A(\vec{k},\omega)$ at the $\vec{k}$
points
outside the Fermi surface possesses 
a strong incoherent tail below the chemical potential.
\section{VERTEX CORRECTIONS}
\label{section4}

   Since quasiparticles move coherently on a reduced energy scale $2J$, 
the Fermi energy is quite small for a small doping case, i.e.\, 
almost on the order of the phonon frequency or even less, which signals a
possible breakdown of the standard strong (phonon) coupling
theory. In the present section, the Migdal-type vertex corrections
are studied in the $t-J-g$ model.  Below
$k$ is defined as $(\vec{k},ik_{n})$  where $k_{n}$ is a  
Matsubara frequency. 
Hence a summation over $k$ means the summation over both
momenta $\vec{k}$ and Matsubara frequencies $k_{n}$.
The lowest order vertex corrections to the hole-phonon interaction
in Fig.~\ref{fig114} can be written as $g\Gamma(k,k+q)$, where
\begin{equation}
 \Gamma(k,k+q) = -\frac{1}{\beta}\sum_{k'}G(k')G(k'+q)B(k-k') \; .
                                                         \label{eq:541}
\end{equation}
$B(k-k')$ is the Green's function of the optical phonon.
Using the spectral representation for the hole Green's function 
and converting the summation over Matsubara frequencies
into a contour integration 
leads to 
\begin{eqnarray}
 \Gamma(k,k+q) & = & g^{2}\sum_{\vec{k}'}\int\int d\omega d\omega'
                 A(\vec{k}',\omega)A(\vec{k}'+\vec{q},\omega')
                 \frac{1}{\omega-\omega'} \hspace{3.0cm}
                                                        \nonumber  \\
        & \times & \{  \frac{1}{-ik_{n}+\omega+\Omega}
                         [F(\omega)-N(\Omega)-1]
                         - \frac{1}{-ik_{n}+\omega-\Omega}
                         [F(\omega)+N(\Omega)]   \hspace{1.0cm}
                                                        \nonumber  \\
                         & - & \frac{1}{-ik_{n}+\omega'+\Omega}
                         [F(\omega')-N(\Omega)-1]
                         + \frac{1}{-ik_{n}+\omega'-\Omega}
                         [F(\omega')+N(\Omega)]  \} \; ,
                                                         \label{eq:543}
\end{eqnarray}
where $F(\omega)$ and $N(\omega)$ are the Fermi and Bose distribution functions
respectively, and the $iq_{n} \rightarrow 0$ limit is taken first for 
numerical simplicity. Since the numerical computation for 
$q \ne 0$ shows a similar result to $q=0$ case, we restrict 
ourselves to the latter case.
By taking the $T=0$ and $\vec{q} \rightarrow 0$ limits and 
the analytic continuation $ik_{n} \rightarrow k_{0}+i\delta$ as well as
by noting that
$\Theta(x)=1-\Theta(-x)$ and that (an integral is in the principal value
sense)
\[  \int d\omega'\frac{A(\vec{k},\omega')}{\omega
     -\omega'} = ReG(\vec{k},\omega) \; , \]
we arrive at
\begin{eqnarray}
  \Gamma(k_{0}) & = & 2g^{2}\sum_{\vec{k}}\int d\omega 
                 A(\vec{k},\omega)ReG(\vec{k},\omega)  %\hspace{18.0cm}
                                                        \nonumber  \\
              & \times & \{  \frac{\Theta(\omega)}
                            {-k_{0}+\omega+\Omega-i\delta}
                          + \frac{\Theta(-\omega)}
                            {-k_{0}+\omega-\Omega-i\delta}  \}
                                                        \nonumber  \\
                & = & 2g^{2}\sum_{\vec{k}}\int d\omega 
                 A(\vec{k},\omega)ReG(\vec{k},\omega)  
                                                        \nonumber  \\
              &      \times & \frac{1}
                 {-k_{0}+\omega+\Omega sign(\omega)-i\delta} \ .  
                                                         \label{eq:545}
\end{eqnarray}
Therefore, the real and imaginary parts of the lowest order
vertex correction are found as
\begin{eqnarray}
 Re\Gamma(k_{0}) & = & 2g^{2}\sum_{\vec{k}}\int_{-\infty}^{\infty} d\omega 
                 A(\vec{k},\omega)ReG(\vec{k},\omega)   %\hspace{3.0cm}
                                                         \nonumber   \\
              &      \times & \frac{1}
                            {-k_{0}+\omega+\Omega sign(\omega) } \ ,
                                                         \label{eq:546}
\end{eqnarray}
and 
\begin{eqnarray}
   Im\Gamma(k_{0})  = \left\{  \begin{array}{lll}
                 2\pi g^{2}\sum_{\vec{k}}
                 A(\vec{k},k_{0}+\Omega)ReG(\vec{k},k_{0}+\Omega)
                                & \mbox{if $k_{0} < -\Omega$}
                                                   \nonumber      \\
                        0
                                & \mbox{if $-\Omega < k_{0} < \Omega$}
                                                   \nonumber       \\
                 2\pi g^{2}\sum_{\vec{k}}
                 A(\vec{k},k_{0}-\Omega)ReG(\vec{k},k_{0}-\Omega)
                     & \mbox{if $k_{0} > \Omega$} \; . 
                                \end{array}
      \right.   
                                                         \label{eq:547}
\end{eqnarray}

   Fig.~\ref{fig115} and Fig.~\ref{fig116} show the real and 
imaginary parts of the lowest order vertex correction for several
hole-phonon coupling constants, respectively. 
As the coupling strength
increases, the real and imaginary parts of the vertex correction grow. 
In spite of the expectation
that the Migdal approximation may break down due to the small hole band
width $2J$ determined from the $t-J$ model \cite{Ramsak:1992},
the vertex correction is
much smaller than unity for up to $g \sim 2J$.
For a noninteracting electron system, first order vertex correction
has been known to be of the order of $\omega_{D}/E_{F}$
\cite{Migdal:1958}, where $\omega_{D}$ is the Debye frequency and 
$E_{F}$ is the Fermi energy.
But, the adiabatic argument valid for a weakly interacting system
breaks down for 
strongly correlated electrons, since the Fermi velocity is comparable
to or even less than the phonon phase velocity in a considerable part of 
the Brillouin zone.
According to Eq.~\ref{eq:546}, first order vertex correction is 
roughly proportional to the square of the quasiparticle residue.
This indicates a possibility that the significant
renormalization $(0.2-0.3)$ of the quasiparticle residue for a strongly 
correlated electron system makes the vertex correction much reduced,
thereby accounting for the small vertex correction from the hole-phonon
interaction in the $t-J-g$ model.
Hence, the present calculation numerically corroborates
using the noncrossing approximation for the self-energy both in the
hole-magnon and hole-phonon interactions. 
\section{OPTICAL CONDUCTIVITY IN THE $\lowercase{t} - J -
         \lowercase{g}$ MODEL}
\label{section5}

The current operator after Bogoliubov transformations of the spin
wave operators becomes \cite{Kyung:19961}
\begin{equation}
  J_{x}(\vec{q})  =  \frac{2et}{N}\sum_{\vec{k},\vec{p}}
             h_{\vec{k}}h^{+}_{\vec{p}} [   
             C_{\vec{k},\vec{p}}\alpha^{+}_{\vec{k}-\vec{p}-\vec{q}}
           + D_{\vec{k},\vec{p}}\alpha_{-\vec{k}+\vec{p}+\vec{q}} ] \; ,
                                                           \label{eq:616}
\end{equation}
where the bare current vertices $C_{\vec{k},\vec{p}},D_{\vec{k},\vec{p}}$
are defined as
\begin{eqnarray}
  C_{\vec{k},\vec{p}} & = & u_{\vec{k}-\vec{p}}\sin p_{x}
                           +v_{\vec{k}-\vec{p}}\sin k_{x} \; ,
                                                          \nonumber  \\
  D_{\vec{k},\vec{p}} & = & v_{\vec{k}-\vec{p}}\sin p_{x}
                           +u_{\vec{k}-\vec{p}}\sin k_{x} \; .
                                                           \label{eq:617}
\end{eqnarray}
Due to the special nature of the interaction vertices, it was
established that the lowest order contribution to the optical 
conductivity dominates at $\omega \neq 0$ \cite{Kyung:19961}.   
Hence, we consider only the lowest order 
diagrams in the present study.
The lowest order contribution comes from two diagrams in 
Fig.~\ref{fig117} owing to the structure of the current 
operator Eq.~\ref{eq:616}.
In the combined limits of zero temperature $(T \rightarrow 0)$ 
and long wavelength
electromagnetic radiation $(\vec{q} \rightarrow 0)$,
the optical conductivity becomes
\begin{eqnarray}
 \sigma_{1}(q_{0}) = \frac{\pi}{q_{0}}(\frac{2t}{N})^{2}
                 \int_{0}^{q_{0}} d\omega
       \sum_{\vec{k}}A(\vec{k},\omega-q_{0})
                                                           \nonumber  \\
      \times  \sum_{\vec{p}} 
       C_{\vec{k},\vec{p}}^{2}
       A(\vec{p},\omega-E_{\vec{k}-\vec{p}}) \; .
                                                            \label{eq:628}
\end{eqnarray}

   Fig.~\ref{fig118} presents the optical 
conductivity for the $t-J-g$ model, as the hole-phonon coupling 
constant varies.
As expected from the corresponding spectral density function, the 
contribution to the conductivity from the multi-magnon excitations
(``string structure'') decreases,
while strong
absorption appears right above the $2J$ peak. This new peak 
in the absorption comes from
the hole-phonon interaction, as can be seen in the corresponding spectral
density function. Generally, the peak at 
$2J$ peak and higher energy 
incoherent spin wave peaks
are suppressed and broadened in the presence of the 
strong hole-phonon interaction. This may be associated with the  
growth of {\em featureless} spectral weight at the mid-infrared region, when 
the CuO$_{2}$ plane is doped with charge carriers \cite{Uchida}.
\section{CONCLUSION}
\label{section6}

    The influence of the hole-phonon interaction on various physical
quantities was studied within the noncrossing  
approximation for the spin wave and optical phonon interactions on the 
same footing.
As the hole-phonon coupling 
constant $g$ increases, 
the quasiparticle residue is further reduced 
and spin wave peaks in the spectral density function and optical conductivity
are more suppressed.  
Phonon peaks in the spectral density function $A(\vec k,\omega)$, instead, 
grow more pronounced around the quasiparticle pole at a
low energy on the scale of $\Omega$ .
A sharp drop in the hole momentum distribution function is found for 
all the hole-phonon coupling constants we have studied.
The invariance of the volume enclosed by the Fermi surface for all 
the chosen hole-phonon coupling constants   
numerically verifies Luttinger's theorem for doped holes in 
the $t-J-g$ model.
Our numerical estimate of the lowest order vertex corrections to the
hole-phonon coupling vertex $g$ due to the
hole-phonon interaction, gives relatively small values $\ll 1$. 
This means that electron-phonon vertex
corrections are not 
important and Migdal's approximation can be used
in calculating the hole self-energy. The smallness of the effective
hole bandwidth of order $2J$, is compensated by  suppressed quasiparticle
residues. 
Due to the presence of additional phonon induced absorption,   
$2J$ and higher hole-multi-spin wave peaks in optical conductivity
are suppressed and broadened. 
\acknowledgments

    This work has been supported by the Center for
Superconductivity Research
of the University of Maryland at College Park and by NASA Grant 
NAG3-1395. The authors are grateful to P. B. Allen and V. J. Emery for 
discussions.
S. I. Mukhin is grateful to colleagues at the Department of Physics 
of the University
of Maryland for their warm hospitality during his stay at College Park. 
\newpage
\vspace{5.0cm}
\begin{table}
\begin{center}
\begin{tabular}{|c||c|c|c|c|c|c|}    
 $g/J$  & $a_{(0,0)}$ & $a_{(0,\pi/2)}$ & $a_{(\pi/2,\pi/2)}$   
   & $a_{(\pi/2,\pi/2)}^{calc}$ & $\Delta\mu/J$ & $\lambda_{num}$                 \\   \hline\hline
0.0  & 0.103 & 0.408 & 0.349 & 0.349 & -5.378 & 0.000  \\ \hline
0.5  & 0.001 & 0.376 & 0.337 & 0.337 & -5.425 & 0.102  \\ \hline
1.0  & 0.002 & 0.322 & 0.304 & 0.305 & -5.603 & 0.424  \\ \hline
1.5  & 0.003 & 0.258 & 0.266 & 0.274 & -5.934 & 0.894  \\ \hline
2.0  & 0.007 & 0.231 & 0.216 & 0.266 & -6.353 & 1.765  
\end{tabular}
\end{center}
\caption{Quasiparticle residue, energy shift and mass renormalization
         factor due to the hole-phonon interaction.
         The first three columns after the column corresponing to the
         value of $g/J$ are quasiparticle residues at three
         different $\vec{k}$ points and the fourth column is the 
         energy shift due to the hole-magnon and hole-phonon interactions 
         at the $(\pi/2,\pi/2)$ point.
         $a_{(\pi/2,\pi/2)}^{calc}$ is the quasiparticle pole strength 
         at the $(\pi/2,\pi/2)$ point
         obtained by perturbation theory. $\lambda_{num}$ is the numerically 
         calculated renormalization factor due to the hole-phonon 
         interaction.}
\label{table2}
\end{table}
\newpage
\begin{figure}
\caption{Four different diagrams contributing to the self-energy,
         magnon emitting and absorbing processes, and 
         phonon emitting and absorbing ones.}
\label{fig110}
\end{figure}
\begin{figure}
\caption{Spectral density function for four different coupling
         constants $g$ at $\vec{k}=(\pi/2,\pi/2)$ for a      
         24 $\times$ 24 momentum cluster.  
         The figures for $g \neq 0$ are shifted upward by 0.2.
         The doping concentration is  $3.2 \%$. The phonon frequency
         $\Omega$ is equal to 0.5$J$.}
\label{fig111}
\end{figure}
\begin{figure}
\caption{Density of states for the dressed holes for four
         different coupling constants $g$ for a 24 $\times$ 24 momentum
         cluster.   
         The figures for $g \neq 0$ are shifted upward by 0.2.
         The doping concentration is  $3.2 \%$.}
\label{fig112}
\end{figure}
\begin{figure}
\caption{Momentum distribution function for four different
         coupling constants $g$ for a 24 $\times$ 24 momentum cluster. 
         It is scanned from 
         S$=(\pi/2,\pi/2)$ to M$=(\pi,0)$. The  ellipse-like  
         Fermi surface pockets are calculated numerically for 
         the doping concentration  $3.2 \%$.}
\label{fig113}
\end{figure}
\begin{figure}
\caption{Lowest order vertex correction 
         due to hole-phonon interaction.}
\label{fig114}
\end{figure}
\begin{figure}
\caption{Real part of the lowest order vertex correction due to 
         hole-phonon interaction for a 24 $\times$ 24 momentum cluster.
         The figures for $g/J \geq 1.0$ are shifted
         upward by 0.1. 
         The doping concentration is  $3.2 \%$.}
\label{fig115}
\end{figure}
\begin{figure}
\caption{Imaginary part of the lowest order vertex correction due to 
         hole-phonon interaction for a 24 $\times$ 24 momentum cluster.
         The figures for $g/J \geq 1.0$ are shifted
         upward by 0.1.
         The doping concentration is  $3.2 \%$.}
\label{fig116}
\end{figure}
\begin{figure}
\caption{(a) Lowest order diagram for the optical conductivity,
            corresponding to an intermediate magnon emitting process.
         (b) Other lowest order diagram for the optical conductivity.
            At zero temperature, this contribution vanishes, since
            it involves a magnon absorption.}
\label{fig117}
\end{figure}
\begin{figure}
\caption{Optical conductivity calculated by the analytic
         expressions given in the text for four different hole-phonon 
         coupling constants $g$ for a 24 $\times$ 24 momentum cluster. 
      The units are  $2 \pi e^2 \times 10^{2}$ choosen for convinience
         (compare Ref. 9).
         The figures for $g \neq 0$ are shifted upward by 20.
         The doping concentration is $3.2 \% $.}
\label{fig118}
\end{figure}
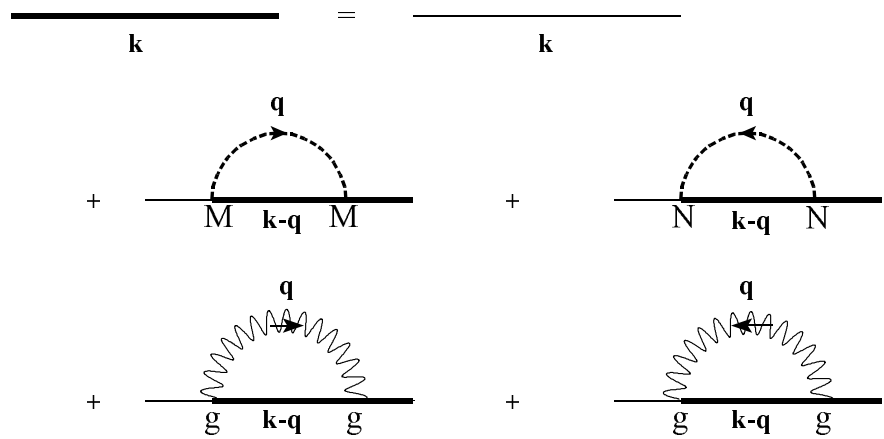
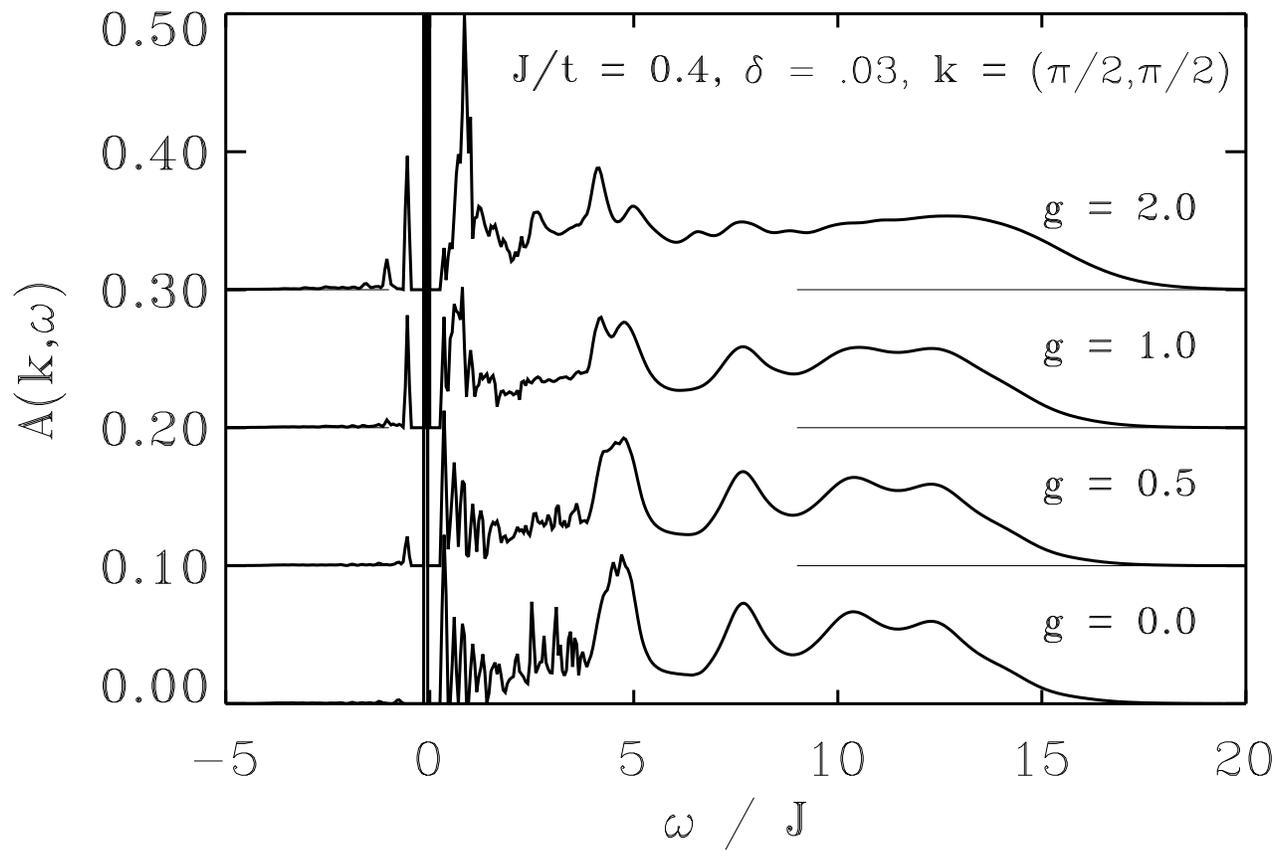
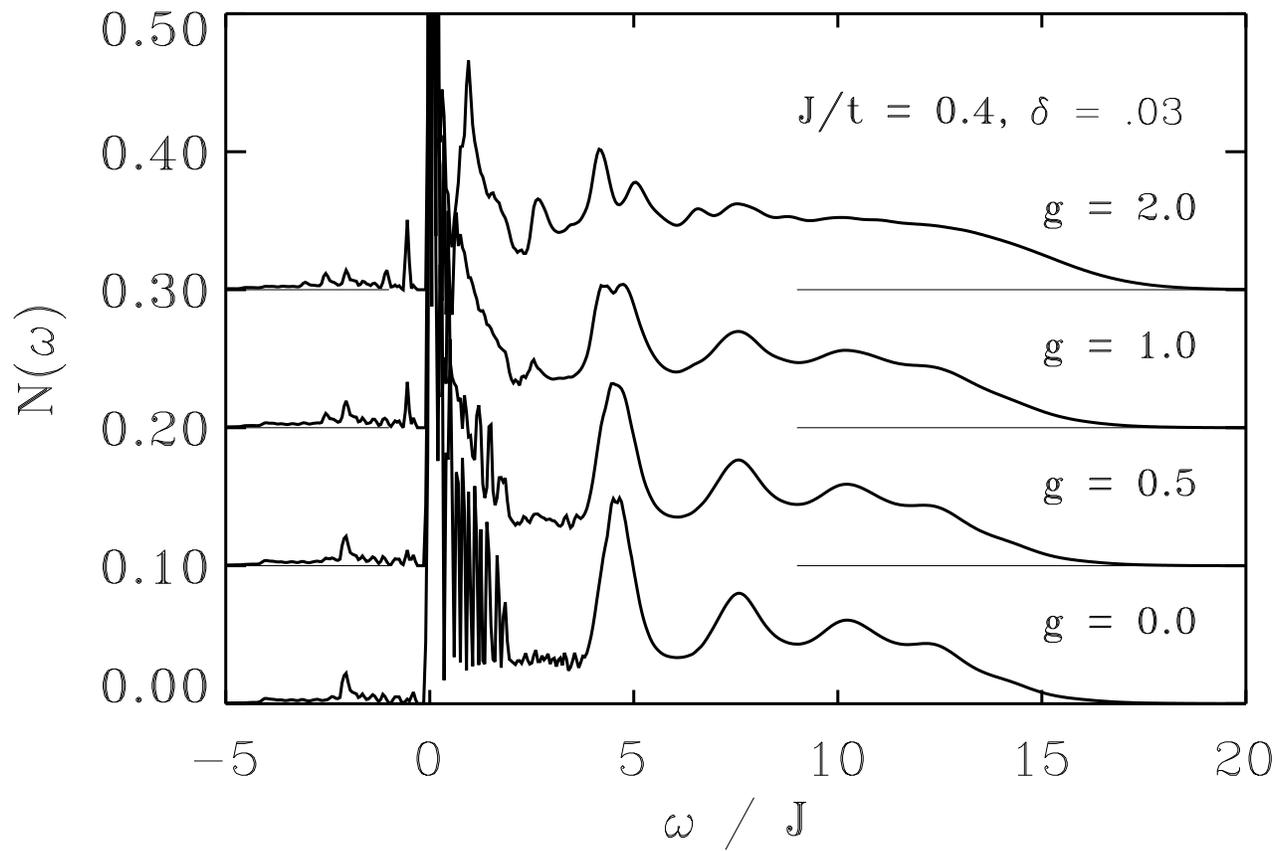
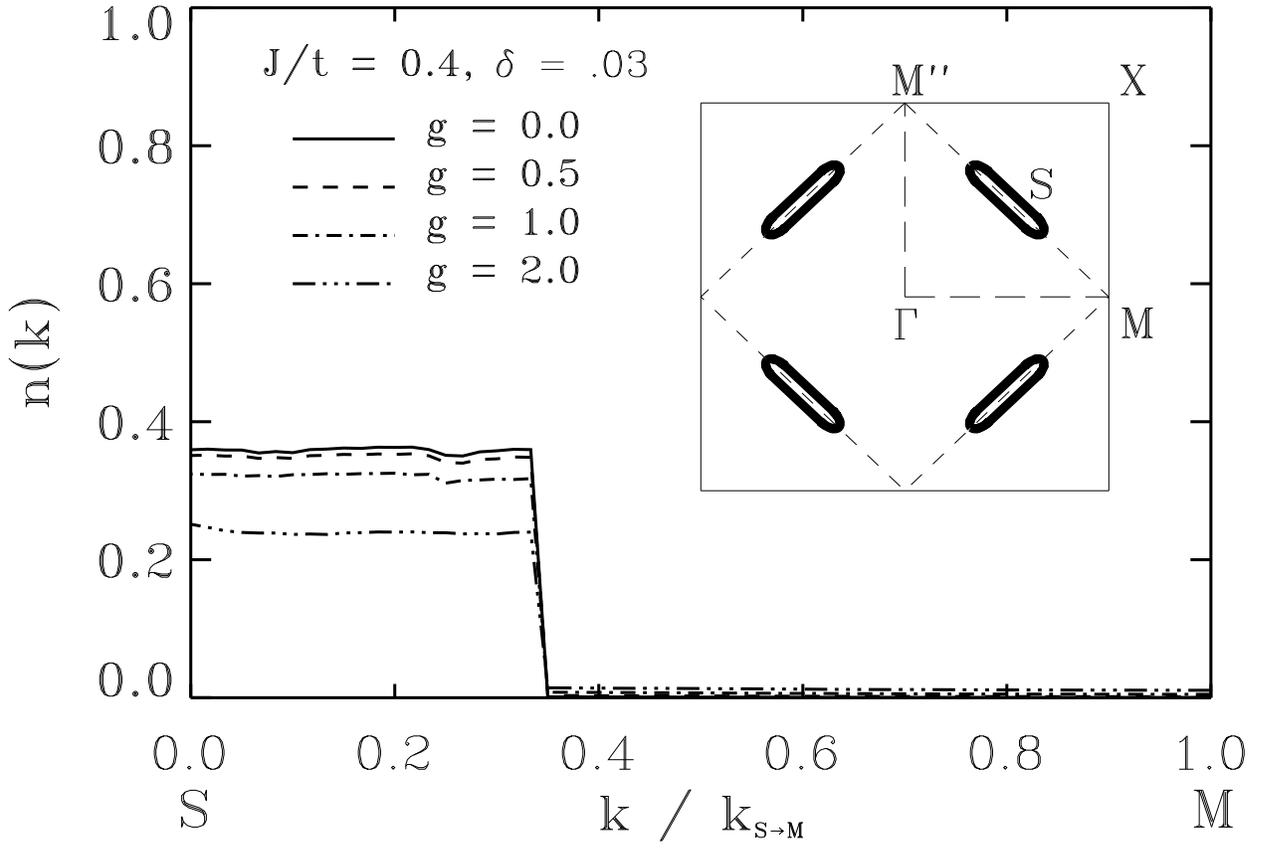
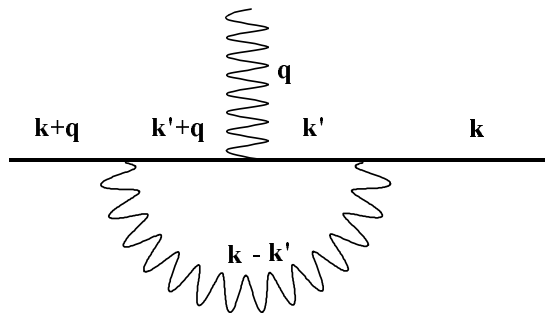
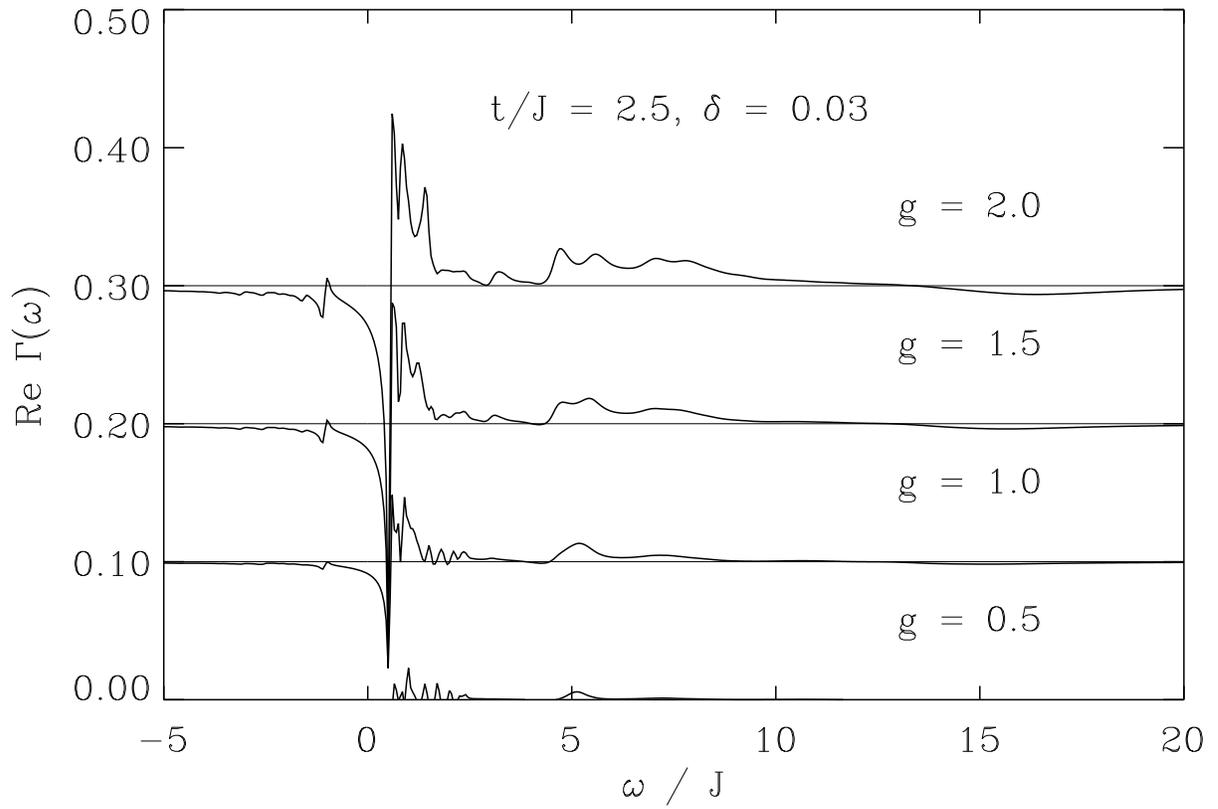
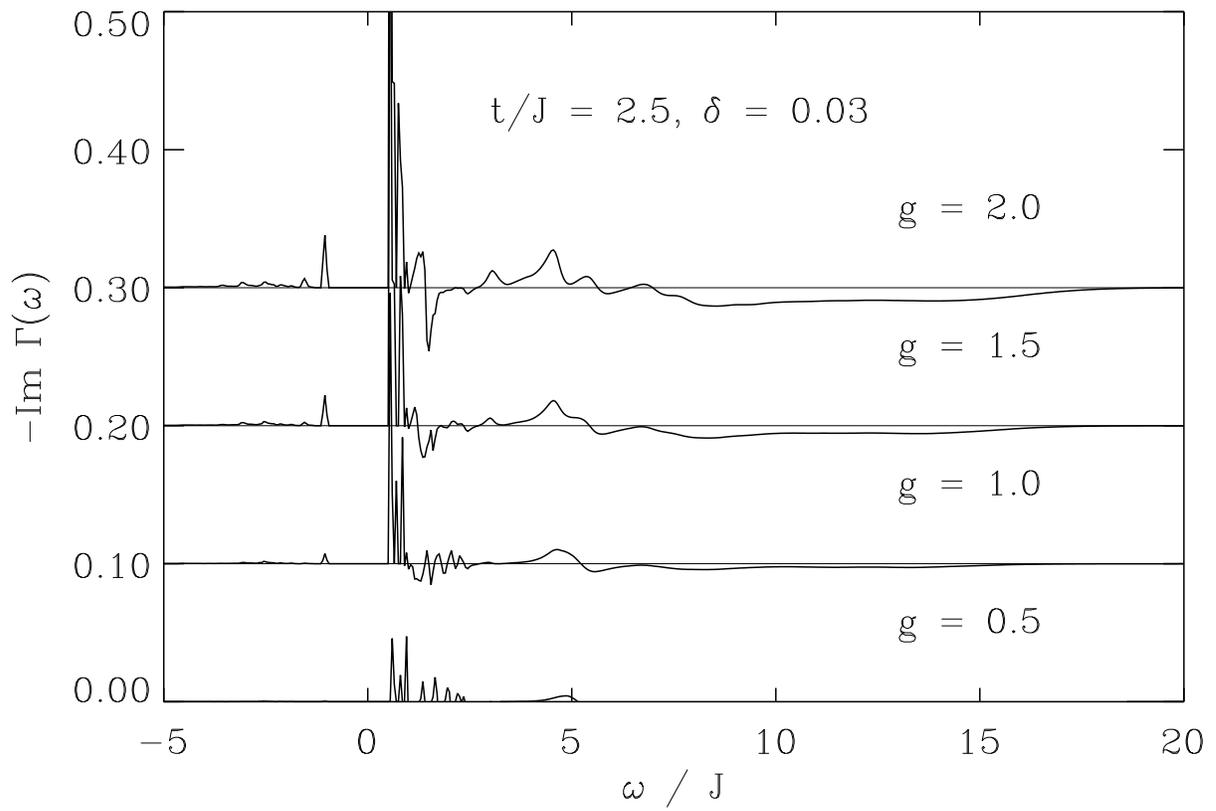
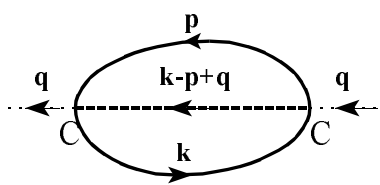
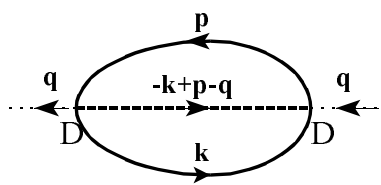
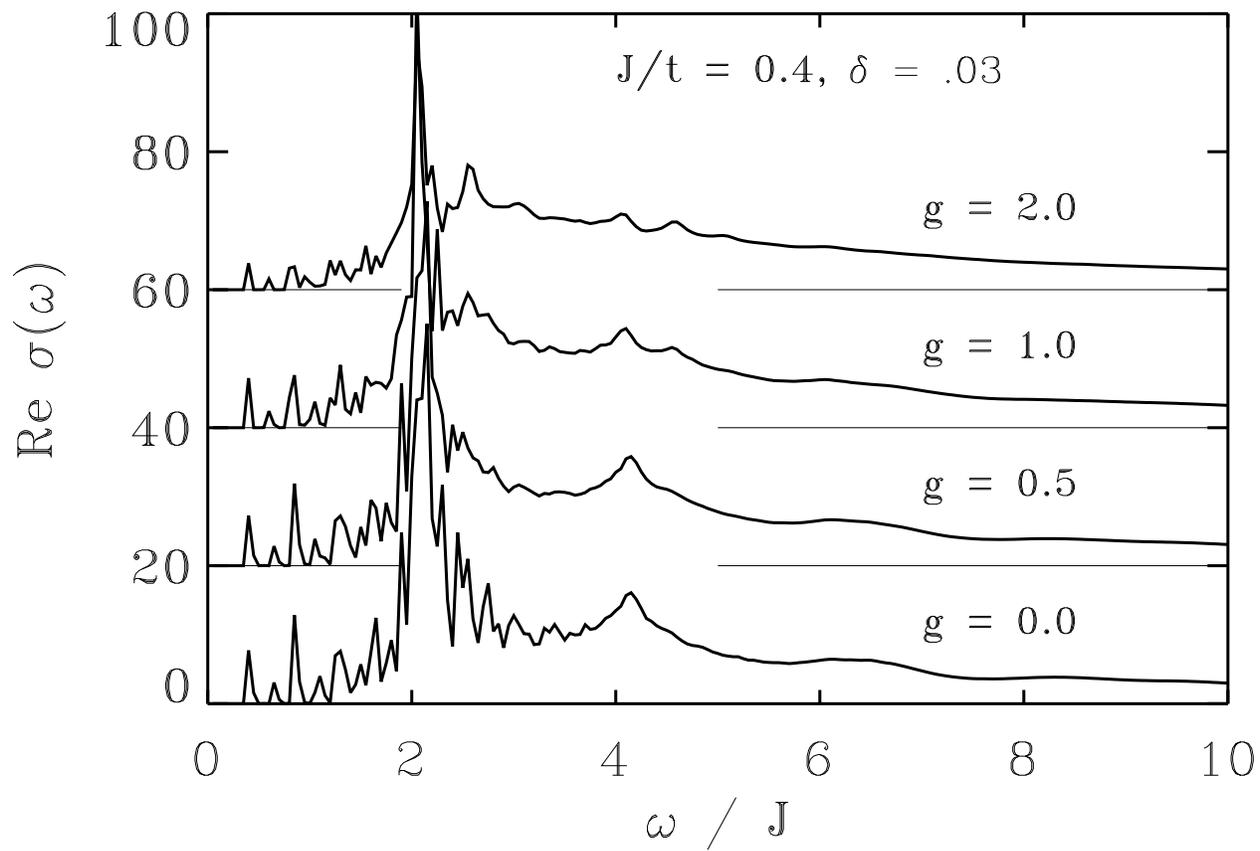
\end{document}